\documentclass[amsmath,aps,pra]{revtex4}
\usepackage[pdftex]{graphicx}
\usepackage{times}
\usepackage{epsfig}
\usepackage{amssymb}
\usepackage{bm}
\usepackage{amsmath}
\usepackage{multirow}

\begin{document}
\title{BBR-induced Stark shifts and level broadening in helium atom}

\author{T. Zalialiutdinov$^{1}$, D. Solovyev$^{1}$,  L. Labzowsky$^{1,2}$}

\affiliation{$^1$ Department of Physics, St.Petersburg State University,
Ulianovskaya 1, Petrodvorets, St.Petersburg 198504, Russia
\\
$^2$  Petersburg Nuclear Physics Institute, 188300, Gatchina, St.
Petersburg, Russia}

\begin{abstract}
The precise calculations of blackbody radiation (BBR)-induced Stark shifts and depopulation rates for low-lying states of helium atom with the use of variational approach are presented. An effect of the BBR-induced induced Stark-mixing of energy levels is considered. It is shown that this effect leads to a significant reduction of lifetimes of helium excited states. As a consequence the influence of Stark-mixing effect on the decay rates of metastable states in helium is discussed in context of formation processes of the cosmic microwave background.
\end{abstract}
\maketitle
\section{Introduction}
The influence of blackbody radiation (BBR) on the atom is the one of important topics of modern atomic physics. Interest to these investigations is stimulated by the essential progress in the theoretical and experimental research of atomic clocks and determination of frequency standards \cite{rilef} that requires a comprehensive analysis of the BBR influence on an atoms \cite{safronova}. 

Recently the temperature-dependent one-loop self-energy (SE) correction of bound atomic electron states was considered in \cite{solovyev} within the frameworks of quantum electrodynamics (QED). According to the results of \cite{solovyev}, the energy shift arises as the real part of self-energy correction, while the imaginary part represents the BBR-induced depopulation rate for a given atomic state. In particular, it was shown that regularization of divergent energy denominators in thermal SE correction leads to an additional contribution to level width. This contribution has not been considered before and can be explained by the level mixing of the atomic states with opposite parity. 

The effect of level-mixing produces a more significant broadening of the emission line then the well-known depopulation rates induced by BBR \cite{farley}. According to \cite{solovyev} the BBR-induced level-mixing leads to a significant reduction of lifetimes even at the room temperatures. The results of evaluation of BBR-induced level-mixing widths show that the width of $2s$ state becomes comparable with the width of $2p$ state at $T\approx 3000$ K. This value is about eight orders of magnitude higher than the natural width, $\Gamma_{2s}=8.229$ $\mathrm{s}^{-1}$. Moreover, in that case the decay of the $2s$ state occurs with the emission of $\mathrm{E1}$ photon. Although this effect is negligible for low temperatures $ T\approx 3 $ K it could be important \cite{timur} for the ionization history of primordial plasma where $2s$ state in hydrogen atom plays the crucial role for formation processes of the Cosmic Microwave Background (CMB) \cite{zeld,peebles}. With the same reasons BBR-induced level mixing effect and corresponding line broadening should be significant for helium atom since its recombination occurs earlier than the hydrogen one and takes place at temperatures $4\times 10^3<T<10^4$ K \cite{seager2000, wongphd}.

Another reason for the further investigation of BBR influence on atoms is the recent accurate measurements of light atoms by the spectroscopic methods. Except hydrogen the most promising atomic system with a sufficiently accurate theory is the helium atom.  The advanced precise methods for the solution of three-body atomic problem were developed in \cite{korobov1,drakebook,kor_monk}. The accuracy of laser spectroscopy experiments of the fine-structure splitting in the $2^3P_J$ levels of ${}^4$He \cite{zheng} reaches 0.13 kHz that allows for the independent determination of the fine-structure constant $\alpha$ with a precision of $2\times 10^{-9}$. Moreover, during the last decades accuracy of experiments with helium has reached the level when the determination of nuclear parameters is possible and, in particular, the determination of nuclear charge radius \cite{yer1,lu}.

In the present paper we calculate  BBR-induced Stark shifts, depopulation rates and BBR-induced level mixing for the helium atom within a variational approach developed in \cite{korobov1}. The paper is organized as follows. The theory of BBR-induced Stark-shift is presented in section \ref{stark}. A contribution to the line-broadening corresponding to the BBR induced level mixing is considered in section \ref{mixing}. In section \ref{var} the brief overview of variational approach and computational details are considered. Section \ref{end} is devoted to the conclusions and discussion of the mixing effect in helium atom to the recombination processes in early universe. Relativistic units ($ \hbar = c = m =1 $) are used throughout this paper unless otherwise stated.

\section{Stark-shift and depopulation rates}
\label{stark}
To describe effects induced by the BBR the quantum mechanical (QM) approach \cite{farley,gallagher}  is usually applied. The extensive use of nonrelativistic QED theory was employed in \cite{escobedo1,escobedo2} for the search of the finite-temperature effects in bound states. The rigorous quantum electrodynamics was applied to derive the Stark shift and widths of the atomic energy levels induced by the BBR in \cite{solovyev}, where perfect agreement between QED (in nonrelativistic limit) and QM results was also demonstrated. The QED description of BBR-induced Stark shift has shown that besides the well-known ac-Stark shift and depopulation rates the quadratic mixing effect of states with opposite parity also arises in the mean BBR electric field \cite{solovyev}. The latter can be estimated with the root mean squared (rms) value (in a.u.)
\begin{eqnarray}
\label{1i}
\langle E^2(\omega) \rangle = \frac{8\alpha^3}{\pi}\frac{\omega^3}{e^{\beta\omega}-1},
\end{eqnarray}
where $\beta = 1/k_B T$, $k_B$ is the Boltzman's constant, $T$ is the radiation temperature in Kelvin, $\alpha$ is the fine structure constant and $ \omega $ is the field frequency. Then the averaged over frequency rms value of electric field is
\begin{eqnarray}
\label{2i}
\langle E^2 \rangle = \frac{1}{2}\int\limits_0^\infty \langle E^2(\omega) \rangle d\omega = \frac{4\pi^3}{15}\alpha^3 (k_B T)^4 =\\\nonumber (8.319\;\mathrm{V/cm})^2\left[T(\mathrm{K})/300\right]^4.
\end{eqnarray}
Following to \cite{farley, solovyev} the BBR-induced Stark-shift for the atomic state of the two-electron atom $a\equiv\lbrace n_a L_a S_a J_a M_a \rbrace$ ($n$ is the principal quantum number, $L$ is the orbital angular momentum, $ S $ is the total electron spin, $J$ is the total electron angular momentum and $M$ is the $z$ component of $J$; $ LS $ coupling scheme is assumed with notations $ n^{2S+1}L_{J} $, where $ 2S+1 $ is the multiplicity of the term $ LS $ and $ J $ enumerates the fine structure levels) in nonrelativistic limit is 
\begin{eqnarray}
\label{18}
\Delta E^{(2)}_a=\frac{4e^2
}{3\pi}\sum_{i=1,2}\sum_{b}\int_{0}^{\infty}d\omega\frac{\omega^3}{e^{\beta \omega}-1}\frac{\omega_{ab}}{\omega_{ab}^2-\omega^2} \\\nonumber\left|\left\langle n_bL_bS_bJ_bM_b\left|\textbf{r}_i \right|n_aL_aS_aJ_aM_a\right\rangle\ \right|^2 ,
\end{eqnarray}
where $\textbf{r}_i$ is radius-vector of the corresponding electron, $e$ is the electron charge, $\left\langle a\left|...\right|b\right\rangle$ denotes the matrix element with Schr\"odinger wave-functions and $\omega_{ab}=E(n_aS_aL_aJ_a)-E(n_bS_bL_bJ_b)$. Expression (\ref{18}) is written in a general case but in nonrelativistic limit within $LS$ coupling scheme and can be easily applied to arbitrary multielectron atom with a substitution of corresponding wave functions \cite{farley}.

The angular integration in the matrix elements in Eq. (\ref{18}) can be performed using standard angular techniques \cite{varsh}
\begin{eqnarray}
\label{wick}
\langle n_bL_bS_bJ_bM_b\left| r^{q} \right| n_aL_aS_aJ_aM_a\rangle=(-1)^{J_b-M_b}\times\\\nonumber
\begin{pmatrix}
J_b & 1 & J_a\\
M_b  & q  & M_a
\end{pmatrix}
\langle n_bL_bS_bJ_b\left|\left| r \right|\right| n_aL_aS_aJ_a\rangle,
\end{eqnarray}
\begin{eqnarray}
\label{wick2}
\langle n_bL_bS_bJ_b\left|\left| r \right|\right| n_aL_aS_aJ_a\rangle=\delta_{S_aS_b}(2J_a+1)^{1/2}\times\\\nonumber
(2J_b+1)^{1/2}
\begin{Bmatrix}
L_b & J_b & S_b\\
J_a  & L_a  & 1
\end{Bmatrix}
\langle n_bL_bS_b \left|\left|r\right|\right| n_aL_aS_a\rangle.
\end{eqnarray}
where $r^q$ is a spherical component of a radius-vector and $\left\langle b\left|\left|...\right|\right|a\right\rangle$ denotes the reduced matrix element. 

A further simplification in Eq. (\ref{18}) follows when we can neglect the fine structure intervals compared to the value $ k_BT $. Then we can ignore the dependence of $ \omega_{ab} $ on $ J_{b} $ and sum over $ J_{b} $ in Eq. (\ref{18}), using equality
\begin{eqnarray}
\label{sumrule}
\sum_{J_b}(2J_b+1)
\begin{Bmatrix}
L_b & J_b & S_b\\
J_a  & L_a  & 1
\end{Bmatrix}
^2
=\frac{1}{2L_{a}+1},
\end{eqnarray}
and substituting Eqs. (\ref{wick})-(\ref{sumrule}) into Eq. (\ref{18}) we obtain
\begin{eqnarray}
\label{swhift}
\Delta E^{(2)}_a=\frac{4e^2
}{3\pi}\frac{1}{2L_{a}+1}\sum_{i=1,2}\sum_{b}
\\
\nonumber
\times \int_{0}^{\infty}d\omega\frac{\omega^3}{e^{\beta \omega}-1}\frac{\omega_{ab}}{\omega_{ab}^2-\omega^2}\left|\langle n_bL_bS_b\left|\left| r_{i} \right|\right| n_aL_aS_a\rangle \right|^2.
\end{eqnarray}

According to \cite{farley}, the BBR efficiently redistributes population among excited states, shortens the atomic lifetimes, and causes the corresponding line broadening. The effective level width is
\begin{eqnarray}
\label{19}
\Gamma^{\mathrm{eff}}_{a}=\Gamma_{a}+\Gamma^{\mathrm{BBR}}_{a},
\end{eqnarray}
where $\Gamma_{a}$ is the natural decay width and $\Gamma^{\mathrm{BBR}}_{a}$ is the BBR-induced width. The natural level width of two-electron atomic state $a$ is given by 
\begin{eqnarray}
\label{radwidth}
\Gamma_{a}=\sum_{n\geqslant 1}\sum_{E_{b}<E_{a}}W^{(n\gamma)}_{ab}.
\end{eqnarray} 
Here summation extends over all states with $E_{b}<E_{a}$ and $W^{(n\gamma)}_{ab}$ is the decay rate for the transition $a\rightarrow b +n\gamma$, where $ n $ is the number of emitted photons \cite{landau}. In the 'length' form one-photon transition rate is (within the dipole approximation)
\begin{eqnarray}
W^{\rm E1}_{ab}=\frac{4e^2
}{3}\frac{\omega_{ab}^3}{2L_{a}+1}\sum_{i=1,2}\left|\langle n_bL_bS_b \left|\left| r_{i} \right|\right| n_aL_aS_a \rangle\right|^2.
\end{eqnarray}
The corresponding equation in 'velocity' form can be obtained with the relation $\langle a \left|\textbf{p}\right| b\rangle=\mathrm{i}\omega_{ab}\langle a \left|\textbf{r}\right| b\rangle$. 

The derivation of the  $ \Gamma^{\mathrm{BBR}}_{a} $ with QM approach was considered in \cite{farley} and in dipole approximation the final result is
\begin{eqnarray}
\label{20}
\Gamma^{\mathrm{BBR}}_{a}=\frac{4e^2
}{3}\frac{1}{2L_{a}+1}\sum_{i=1,2}\sum_{b}\frac{\omega_{ab}^3}{e^{\beta\omega_{ab}}-1}
\\
\nonumber
\times\left|\left\langle n_bL_bS_b\left|\left| r_{i} \right|\right| n_aL_aS_a\right\rangle\ \right|^2 .
\end{eqnarray}
The corresponding partial width $ \Gamma^{\mathrm{BBR}}_{aa'} $ connected with the transition to the state $ b=a' $ is
\begin{eqnarray}
\label{20a}
\Gamma^{\mathrm{BBR}}_{aa'}=\frac{4e^2
}{3}\frac{\omega_{aa'}^3}{2L_{a}+1}\sum_{i=1,2}\frac{\omega_{aa'}^3}{e^{\beta\omega_{aa'}}-1}\\\nonumber\times\left|\left\langle n_{a'}L_{a'}S_{a'}\left|\left| r_{i} \right|\right| n_aL_aS_a\right\rangle\ \right|^2.\qquad
\end{eqnarray}
Expression (\ref{20a}) represents the BBR-induced decay rate if $E_{a'}<E_{a}$ and absorption rate if $E_{a'}>E_{a}$.

\begin{table}[ht]
\caption{The spontaneous decay rates for transitions $a\rightarrow a'+n\gamma(\mathrm{E1})$, where $n$ is the number of emitted photons. All values are given in s$^{-1}$. The number in parentheses indicates the power of ten.}
\begin{tabular}{  c   c   c  }
\hline
\hline
$a-a'$ & $ n $ & $ W^{(n\gamma)}_{aa'} $ \\
\hline
$ 2^1S-1^1S $ & $ 2 $ & $ 51.02 $ \cite{derivianko}\\
$ 2^3S-1^1S $ & $ 2 $ & $ 3.17(-9) $ \cite{derivianko}\\
$ 2^1P-1^1S $ & $ 1 $ & $ 1.79892(9) $ \\
$ 2^1P-2^1S $ & $ 1 $ & $ 1.97458(6) $ \\
$ 2^3P-1^1S $ & $ 1 $ & $ 177.61 $ \cite{drake_tab}\\
$ 2^3P-2^3S $ & $ 1 $ & $ 1.02162(7) $ \\
\hline
\end{tabular}
\label{rate}
\end{table}

\begin{table}[ht]
\begin{center}
\caption{Nonrelativistic energies of helium states obtained in the present work in a.u. Calculations were performed by variational method \cite{korobov1} with a basis set length $ N=500 $.}
\label{energy}
\begin{tabular}{ c  c  c }
\hline
\hline
State & Value obtained in this work & Drake \cite{drakebook} \\
\hline
$ 1^1S $ & $ -2.9037243770  $  & $ -2.9037243770341195 $ \\
$ 2^1S $ & $ -2.1459740460  $  & $ -2.145974046054419  $ \\
$ 2^3S $ & $ -2.1752293782  $  & $ -2.1752293782367913 $ \\
$ 2^1P $ & $ -2.1238430864  $  & $ -2.123843086498093  $ \\
$ 2^3P $ & $ -2.1331641908  $  & $  -2.133164190779273 $ \\
\hline
\hline
\end{tabular}
\end{center}
\end{table}

\begin{table*}[ht]
\caption{The BBR-induced dynamic Stark shifts (in Hz) of energy levels of helium at different temperatures $T$. The first line in each column represents the values calculated with the length of intermediate states basis $N=150$, while the second one corresponds to values calculated with $N=300$. In the second column, the lower line for each state indicates the results obtained in \cite{farley}.}
\label{stark_tab}
\begin{tabular}{  c   c   c   c   c   c  }
\hline
\hline
State & $ T=300 $ K & $ T=1000 $ K & $ T=3000 $ K & $ T=5000 $ K & $ T=10^4 $ K\\
\hline
$ 1^1S $ &  $ -0.0118308 $ & $ -1.47020 $ & $ -119.288 $ & $ -923.580 $ & $ -15021.1 $ \\
$      $ &  $ -0.0118519 $ & $ -1.47017 $ & $ -119.334 $ & $ -923.941 $ & $ -15032.9 $ \\
$      $ &  $ -0.16      $ & $          $ & $  $ & $  $ & $  $ \\
$ 2^1S $ &  $ -7.14049 $ & $ -1186.50 $ & $ 3133.29 $ & $ 78224.9 $ & $ 620851 $ \\
$      $ &  $ -7.14156 $ & $ -1186.62 $ & $ 3125.79 $ & $ 78148.0 $ & $ 619594 $ \\
$      $ &  $ -7.6     $ & $          $ & $  $ & $  $ & $  $ \\
$ 2^1P $ &  $ -0.971738$ & $ -386.414 $ & $ -88444.6 $ & $ -381646 $ & $ -384362 $ \\
$      $ &  $ -0.971905$ & $ -386.417 $ & $ -88444.9 $ & $ -381642 $ & $ -384398 $ \\
$      $ &  $  0.738   $ & $          $ & $           $ & $        $ & $  $ \\
$ 2^3P $ &  $ -1.50882 $ & $ -179021 $ & $ -21010.9 $ & $ -174346 $ & $ -293213 $ \\
$      $ &  $ -1.50882 $ & $ -179042 $ & $ -21307.8 $ & $ -176202 $ & $ -293232 $ \\
$      $ &  $ -0.273   $ & $         $ & $  $ & $  $ & $  $ \\
$ 2^3S $ &  $ -2.77836 $ & $ -388.351 $ & $ -23907.9 $ & $ -16566.6 $ & $ 510.397 $ \\
$      $ &  $ -2.74364 $ & $ -384.020 $ & $ -23508.2 $ & $ -12612.9 $ & $ 545.448 $ \\
$      $ &  $ -3.15    $ & $  $ & $  $ & $  $ & $  $ \\
\hline
\hline
\end{tabular}
\end{table*}

\begin{table*}[ht]
\caption{The values of depopulation rates $\Gamma^{\mathrm{BBR}}_{a}$ (in s$^{-1}$) at different temperatures $T$. The 'length' form corresponds to the first subline, while the 'velocity' form corresponds to the second one. The third subline in the second column indicates the results obtained in \cite{farley}. The number in parentheses indicates the power of ten. Calculations were performed with the intermediate basis length $N=300$. }
\label{depop}
\begin{tabular}{  c   c   c   c   c   c   c  }
\hline
\hline
State & $ T=300 $ K & $ T=1000 $ K & $ T=3000 $ K & $ T=5000 $ K & $ T=10^4 $ K & $ \Gamma_{a} $\\
\hline
$ 2^1S $ &  $ 0.000452739 $ & $ 5469.89 $ & $ 641844 $ & $ 2.10029(6) $ & $ 9.91913(6) $ & $ 51.02 $ \cite{derivianko} \\
$      $ &  $ 0.000452828 $ & $ 5470.97 $ & $ 641970 $ & $ 2.10067(6) $ & $ 9.91973(6) $ & $  $ \\
$      $ &  $ 0.0001      $ & $         $ & $        $ & $            $ & $            $ & $  $ \\
$ 2^3S $ &  $ 1.80706(-12)$ & $ 52.1895 $ & $ 370546 $ & $ 2.33433(6) $ & $ 1.23342(7) $ & $ 3.17(-9) $ \cite{derivianko} \\
$      $ &  $ 1.80697(-12)$ & $ 52.1869 $ & $ 370528 $ & $ 2.33420(6) $ & $ 1.23331(7) $ & $  $ \\
$      $ &  $ 6(-11)      $ & $         $ & $        $ & $            $ & $            $ & $  $ \\
$ 2^1P $ &  $ 1.50913(-4) $ & $ 1823.40 $ & $ 388130 $ & $ 3.96617(6) $ & $ 3.90781(7) $ & $ 1.80089(9) $ \\
$      $ &  $ 1.50913(-4) $ & $ 1823.41 $ & $ 388908 $ & $ 3.98178(6) $ & $ 3.92800(7) $ & $  $ \\
$      $ &  $ 4(-4)       $ & $         $ & $        $ & $            $ & $            $ & $  $ \\
$ 2^3P $ &  $ 6.0252(-13) $ & $ 17.4167 $ & $ 203731 $ & $ 2.92180(6) $ & $ 3.54150(7) $ & $ 1.02164(7) $ \\
$      $ &  $ 6.0252(-13) $ & $ 17.4167 $ & $ 203872 $ & $ 2.92598(6) $ & $ 3.54858(7) $ & $  $  \\
$      $ &  $ 2(-11)      $ & $         $ & $        $ & $            $ & $            $ & $  $  \\
\hline
\hline
\end{tabular}
\end{table*}

\begin{table*}[ht]
\caption{The partial widths $\Gamma^{\mathrm{BBR}}_{aa'}$ (in s$^{-1}$) of helium energy levels at different temperatures $T$. The 'length' form corresponds to the first subline, while the 'velocity' form corresponds to the second one. The number in parentheses indicates the power of ten.}
\label{depoppart}
\begin{tabular}{  c   c   c   c   c   c  c  }
\hline
\hline
State & $ T=300 $ K & $ T=1000 $ K & $ T=3000 $ K & $ T=5000 $ K & $ T=10^4 $ K \\
\hline
$ 2^1S $ & $ 0.000452739 $ & $ 5469.89 $ & $ 638864 $ & $ 1.94492(6) $ & $ 5.85697(6) $ \\
$      $ & $ 0.000452828 $ & $ 5470.97 $ & $ 638990 $ & $ 1.94530(6) $ & $ 5.85813(6) $ \\
$ 2^3S $ & $ 1.80706(-12) $ & $ 52.1895 $ & $ 370414 $ & $ 2.31340(6) $ & $ 1.10458(7) $  \\
$      $ & $ 1.80697(-12) $ & $ 52.1869 $ & $ 370396 $ & $ 2.31329(6) $ & $ 1.10453(7) $  \\
$ 2^1P $ & $ 1.37982(-23) $ & $ 0.09354 $ & $ 161741 $ & $ 2.89645(6) $ & $ 2.78705(7) $ & \\
$      $ & $ 1.38615(-23) $ & $ 0.09397 $ & $ 162483 $ & $ 2.90974(6) $ & $ 2.79984(7) $ & \\
$ 2^3P $ & $ 6.0252(-13) $ & $ 17.4033 $ & $ 190841 $ & $ 2.54616(6) $ & $ 2.59884(7) $ & \\
$      $ & $ 6.0252(-13) $ & $ 17.4033 $ & $ 190976 $ & $ 2.54972(6) $ & $ 2.60332(7) $ & \\
\hline
\hline
\end{tabular}
\end{table*}

\section{BBR-induced level mixing}
\label{mixing}
It is known that the external electric field leads to the Stark mixing of states with opposite parity \cite{Ans,sslg}. This effect is most pronounced for close-lying states ($2s$ and $2p$ states in hydrogen, for example). As it was shown in \cite{solovyev} QED derivation of the BBR-induced Stark-shift and decay rates requires an accurate regularization of divergent energy denominators. Then the Stark shift modifies slightly and includes the Lamb shift. The most interesting result arises with the account for imaginary part of energy denominators (level widths). In this case the BBR-induced mixing effect for the states with opposite parity can be obtained. We should note that this effect can be rigorously derived within the QED theory only (within QM approach the level widths are regarded phenomenologically). 

Derivation and detailed analysis of level mixing effect induced by the BBR were made in general case in \cite{solovyev} and can be applied to the helium atom with the substitution of corresponding wave functions into expression
\begin{widetext}
\begin{eqnarray}
\label{21}
\Gamma_a^{\mathrm{mix}} = \frac{2e^2}{3\pi}\frac{1}{2L_a+1}\sum_{i=1,2}\sum\limits_{b}
|\langle n_bL_bS_b\left|\left| r_i \right|\right| n_aL_aS_a \rangle|^2\\\nonumber
\int\limits_{0}^{\infty}d\omega n_\beta(\omega) \omega^3
\left[\frac{\Gamma_{ba}}{(\tilde{\omega}_{ba}+\omega)^2 + \frac{1}{4}\Gamma_{ba}^2} + \frac{\Gamma_{ba}}{(\tilde{\omega}_{ba}-\omega)^2 + \frac{1}{4}\Gamma_{ba}^2}\right],
\end{eqnarray}
\end{widetext}
where
\begin{eqnarray}
\label{nbeta}
n_\beta(\omega)=\frac{1}{e^{\beta\omega}-1},
\end{eqnarray}
and $\tilde{\omega}_{ba}\equiv E_{b}-E_{a}+\Delta E^{L}_{ba}$, $\Delta E^{L}_{ba}$ is the corresponding Lamb shift, $\Gamma_{ba}\equiv \Gamma_b+\Gamma_a$. Summation in Eq. (\ref{21}) is extended over all states of parity opposite to the parity of state $a$. 

In \cite{solovyev} it was noted that the most intriguing result arises for the metastable $2s$ state in hydrogen atom. In particular, the level mixing effect leads to the additional one-photon electric dipole emission channel of this state. Then the corresponding magnitude of level width exceeds significantly natural level width and BBR-induced depopulation rate even at the room temperatures. The same situation arises for the $2^1S$ and $2^3S$ states in helium which decay via two-photon transitions $2^1S\rightarrow 1^1S+2\gamma(\mathrm{E1})$ and $2^3S\rightarrow 1^1S+2\gamma(\mathrm{E1})$, respectively. The last one is allowed only due to the spin-orbit interaction and was considered in \cite{bely, drake1, derivianko}. Numerical values of two-photon transition rates $2^1S\rightarrow 1^1S+2\gamma(\mathrm{E1})$ and $2^3S\rightarrow 1^1S+2\gamma(\mathrm{E1})$ in absence of external fields are given in Table~\ref{rate}. 

The partial width $\Gamma_{aa'}^{\mathrm{mix}}$ can be introduced as
\begin{widetext}
\begin{eqnarray}
\label{part}
\Gamma_{aa'}^{\mathrm{mix}} = \frac{2e^2}{3\pi}\frac{1}{2L_a+1}\sum_{i=1,2}
|\langle n_{a'}L_{a'}S_{a'}\left|\left|r_i \right|\right| n_aL_aS_a \rangle|^2\\\nonumber
\int\limits_{0}^{\infty}d\omega n_\beta(\omega) \omega^3
\left[\frac{\Gamma_{a'a}}{(\tilde{\omega}_{a'a}+\omega)^2 + \frac{1}{4}\Gamma_{a'a}^2} + \frac{\Gamma_{a'a}}{(\tilde{\omega}_{a'a}-\omega)^2 + \frac{1}{4}\Gamma_{a'a}^2}\right],
\end{eqnarray}
\end{widetext}
where $ a' $ is a first nonvanishing term in the sum over $ b $ in Eq. (\ref{21}). Of particular interest of Eq. (\ref{part}) is in the case when $ a=2^{1(3)}S $ and $ a'=2^{1(3)}P $, i.e. partial width $ \Gamma_{2^{1(3)}S, 2^{1(3)}P\rightarrow 1^1S}^{\mathrm{mix}} $. The latter represents the one-photon decay of the mixed  $ \overline{2^{1(3)}S} $ state \cite{solovyev,timur}. It is important to note that the frequencies of photons emitted in transitions $ \overline{2^1S}\rightarrow 1^1S+2\gamma(\mathrm{E1)} $ and $ \overline{2^3S}\rightarrow 1^1S+2\gamma(\mathrm{E1)} $ are $ \omega_{2^1S,1^1S}=E_{2^1S}-E_{1^1S} $ and $ \omega_{2^3S,1^1S}=E_{2^3S}-E_{1^1S} $ respectively. 
\begin{table*}
\caption{The BBR-induced level-mixing widths $\Gamma^{\mathrm{mix}}_{a}$ (in s$^{-1}$) of helium energy levels at different temperatures $T$. Calculations were performed with the intermediate basis length $N=300$. }
\label{mixing_tab}
\begin{tabular}{  c   c   c   c   c   c  }
\hline
\hline
State & $ T=300 $ K & $ T=1000 $ K & $ T=3000 $ K & $ T=5000 $ K & $ T=10^4 $ K\\
\hline
$ 2^1S $ &  $ 238.879 $  & $ 2.18242(7) $ & $ 2.54447(9) $ & $ 7.81302(9)  $ & $ 2.71061(10) $ \\
$ 2^3S $ &  $ 0.842754 $ & $ 1917.88    $ & $ 1.27973(7) $ & $ 8.08185(7)  $ & $ 1.35344(9)  $ \\
$ 2^1P $ &  $ 112.408  $ & $ 7.27742(6) $ & $ 4.53458(9) $ & $ 7.01676(10) $ & $ 9.9988(11)  $ \\
$ 2^3P $ &  $ 0.995895 $ & $ 795.448    $ & $ 9.17896(8) $ & $ 2.4130(10)  $ & $ 3.04126(11) $ \\
\hline
\hline
\end{tabular}
\label{gmix}
\end{table*}
\begin{table*}
\caption{The partial widths $\Gamma^{\mathrm{mix}}_{aa'}$ (in s$^{-1}$) of energy levels of helium at different temperatures $T$. }
\label{mix-part}
\begin{tabular}{  c   c   c   c   c   c  }
\hline
\hline
State & $ T=300 $ K & $ T=1000 $ K & $ T=3000 $ K & $ T=5000 $ K & $ T=10^4 $ K\\
\hline
$ 2^1S $ &  $ 237.873  $ & $ 2.18241(7) $ & $ 2.54316(9)    $ & $ 7.74522(9) $ & $ 2.33423(10) $ \\
$ 2^3S $ &  $ 0.256627 $ & $ 1845.36    $ & $ 1.27869(7)    $ & $ 7.98911(7) $ & $ 3.81636(8) $ \\
$ 2^1P $ &  $ 28.3746  $ & $ 4239.14    $ & $ 4.64835(8)    $ & $ 8.30987(9) $ & $ 7.99245(10) $ \\
$ 2^3P $ &  $ 0.80590  $ & $ 770.401    $ & $ 9.16819(8)    $ & $ 2.40758(10)$ & $ 3.02379(11) $ \\
\hline
\hline
\end{tabular}
\label{gmixpar}
\end{table*}

\section{Variational approach and computational details}
\label{var}
For the numerical calculations in two-electron atom we use the trial wave functions with quasirandom nonlinear parameters developed in \cite{korobov1,kor_monk}. The wave function with the certain values of electron angular momentum $ L  $, its projection $ M $ and parity $ \pi=(-1)^{L} $ is
\begin{eqnarray}
\label{13}
\Psi_{LM}\left(\textbf{r}_1\textbf{r}_2\right) = \sum\limits_{l_{1}+l_{2} = L}\left[Y^{LM}_{l_{1}l_{2}}\left(\textbf{n}_1,\textbf{n}_2\right)G_{l_{1}l_{2}}^{L\pi}\left(r_1,r_2\right)\qquad
\right. 
\\
\nonumber
 \left. 
\pm (1\leftrightarrow 2)\right],\qquad\qquad\qquad\qquad\qquad
\end{eqnarray}
where $G_{l_{1}l_{2}}^{L\pi}$ is the radial part and $Y^{LM}_{l_{1}l_{2}}$ is the corresponding angular part \cite{varsh}. The sign $+$ or $-$ in Eq. (\ref{13}) is refered to the singlet or triplet state, respectively. Following to procedure \cite{korobov1} the radial part $G_{l_{1}l_{2}}^{L\pi}$ is expanded into exponential basis set with complex coefficients $\alpha_{i}$, $\beta_{i}$ and $\gamma_{i}$  
\begin{eqnarray}
\label{14}
G_{l_{e}l_{e}}^{L\pi}\left(r_1,r_2\right)=\sum\limits_{i=1}^{N}\left\lbrace U_{i}\;\mathrm{Re}\left[e^{-\alpha_{i}r_1-\beta_{i}r_2-\gamma_{i}r_{12}}\right]\right. \\\nonumber +\left. W_{i}\;\mbox{Im}\left[e^{-\alpha_{i}r_1-\beta_{i}r_2-\gamma_{i}r_{12}}\right]\right\rbrace,
\end{eqnarray}
where $r_{12}=\left|\textbf{r}_1-\textbf{r}_2\right|$, $U_{i}$ and $W_{i}$ are the linear parameters requiring optimization. The choice of nonlinear parameters for the helium states is discussed in \cite{korobov1}. Then the reduced matrix elements in Eqs. (\ref{18}), (\ref{20}) and (\ref{21}) with the wave-functions (\ref{13}) can be calculated in a closed analytical form, see \cite{drake}. 

As a first step to test the methods of calculations the nonrelativistic energies of helium states were evaluated, see Table~\ref{energy}, which are in a good agreement with the values given in \cite{drakebook}. Variational parameters for the initial states were optimised to reach ten decimal digits precision in eigenvalues, which is quite enough for calculations of Stark shifts and transition probabilities. For the numerical calculations of Stark shifts, depopulation rates and BBR-induced level mixing the different sets of basis states were employed. All the initial states $a$ were evaluated with the basis length $N=500$. In order to test the convergence of results for the Stark shifts the basis of intermediate states was employed with two different lengths $N=150$ and $N=300$ (see Table~\ref{stark_tab}). Calculations of depopulation rates were performed in the 'length' and 'velocity' forms. This also justifies obtained values. 

\section{Conclusions}
\label{end}
Evaluation of Stark shifts, depopulation rates and BBR-induced level mixing widths of helium states with the use of precise variational wave functions were performed. Results of calculations of dynamic Stark shifts and depopulation rates are in a sufficient agreement with the values presented in \cite{farley}. In \cite{farley} the method of quantum defect was used for the calculations of Stark shifts and depopulation rates. This is more suitable for the evaluation of Rydberg states than low-lying ones. Therefore the variational approach applied the present calculations gives more precise results.

The values of BBR-induced Stark shift in Table~\ref{stark_tab} can be important for the precise determination of transition frequencies. The results for depopulation rates $\Gamma^{\mathrm{BBR}}_{a}$ are presented in Table~\ref{depop}. These values were calculated in the 'length' and 'velocity' forms to check the numerical methods. The partial depopulation widths $\Gamma^{\mathrm{BBR}}_{aa'}$ are given in Table~\ref{depoppart}. These results show that the BBR-induced widths are headed by the corresponding partial decays to the ground state at the room temperatures. The increasing of temperature leads to the more significant role of transition rates to upper states. It is important to note that in the present work we do not consider other radiative corrections that depend on different powers of $T$. This requires a separate study and we leave it for future works.

The BBR-induced level mixing widths $\Gamma^{\mathrm{mix}}_{a}$ and corresponding partial widths $\Gamma^{\mathrm{mix}}_{aa'}$ are given in Tables~\ref{mixing_tab} and \ref{mix-part}, respectively. The comparison of these two magnitudes reveals that the leading contribution to the $\Gamma^{\mathrm{mix}}_{a}$ arises from the decay of mixed state to the ground one. The most important result is that $\Gamma^{\mathrm{mix}}_{a}$ exceeds significantly $\Gamma^{\mathrm{BBR}}_{a}$ at all temperatures, see Tables~\ref{depop} and \ref{mixing_tab}. The reason is an additional one-photon decay channel which is allowed due to the mixing of states with opposite parity.

As it was found in \cite{timur} the effect of BBR-induced level mixing influences significantly on the processes of radiation escape from the matter in cosmological recombination epoch of the early universe. The ionization fraction undergoes modification upto the level of 20$\%$ for the $2s$ state in hydrogen atom with the account for mixing effect. Despite the period of recombination is almost the same, the essential changes of the CMB temperature fluctuations map is expected in the far tail of multipole expansion. In these aspects the helium atom should be considered also \cite{seager2000}. 

The period of helium recombination in the primordial plasma refers to the redshift $1600<z<3000$, where an important role plays the two-photon transition $2^1S \rightarrow 1^1S +2\gamma(\mathrm{E1})$ in helium atom.  Since the CMB has a blackbody spectrum the effect of BBR-induced level-mixing can be taken into account in the same way as described above. The presence of blackbody radiation makes the metastable $2^1S$ state in helium atom to decay with the emission of one-photon electric dipole photon. 
This becomes possible due to the effect of BBR-induced level-mixing \cite{solovyev}. According to the results presented in \cite{solovyev} the BBR-induced electric field leads to the mixing of states with opposite parity in an atom,  $2^1S$  and $2^1P$ states in our case. Consequently the admixed state  $\overline{2^1S}$ decays via electric dipole transition  $\overline{2^1S}\rightarrow 1^1S +\gamma(\mathrm{E1})$. The probability of this process dominates over the probability of spontaneous two-photon transition $2^1S\rightarrow  1^1S +2\gamma(\mathrm{E1})$ even at the room temperatures, see Table~\ref{rate} and \ref{gmix}. With the increasing of temperature to thousands of kelvin the decay rates of $2^1S$ and $2^1P$ states becomes comparable. 

The modern theory of cosmic microwave background \cite{seager2000} was developed without the account for the mixing effect and also without the forbidden two-photon decay of $2^3S$ helium state. The last transition is about ten order less then allowed $2^1S\rightarrow 1^1S+2\gamma(\mathrm{E1})$ decay channel (see Table~\ref{rate}) and occurs via the spin-orbit mixing of states \cite{breit}. However, the results of calculations (second line in Table~\ref{gmixpar}) reveals that in the presence of BBR the $2^3S$ state in helium atom decays due to the admixture of  $2^3P$ state via the E1 transition. The corresponding transition rate exceeds significantly the natural width even at the room temperature. Thus, we conclude that the level mixing in the helium atom induced by the blackbody radiation can affect essential on the ionization history of primordial plasma. The more detailed research on this subject should include the solution of rate equations and we omit this for further investigations.

\section*{Acknowledgements}
This work was supported by Russian Science Foundation (grant 17-12-01035). The authors are indebted to V. I. Korobov for permission to use the {\it Fortran} code for the construction of the He variational wave functions.


\begin{thebibliography}{100}

\bibitem{rilef} F. Riehle, {\it Frequency Standards: Basics and Applications}, Wiley-VCH, (2006).

\bibitem{safronova} M. S. Safronova, M. G. Kozlov, and Charles W. Clark, Phys. Rev. Lett. {\bf 107}, 143006 (2011).

\bibitem{solovyev} D. Solovyev, L. Labzowsky, and G. Plunien, Phys. Rev. A {\bf 92}, 022508 (2015).

\bibitem{farley} J. W. Farley and W. H. Wing, Phys. Rev. A {\bf 23}, 2397 (1981).

\bibitem{timur} T. Zalialiutdinov, D. Solovyev, L. Labzowsky, and G. Plunien, Phys. Rev. A {\bf 96}, 012512 (2017)

\bibitem{zeld} Ya. B. Zel'dovich, V. G. Kurt, and R. A. Sunyaev, Zh. Eksp.
Teor. Fiz. 55, 278 (1968) , [Sov. Phys.–JETP 28, 146 (1969)].

\bibitem{peebles} P. J. E. Peebles, Astrophys. J. 153, 1 (1968).

\bibitem{seager2000} S. Seager, D. D. Sasselov, and D. Scott, ApJS, 128, 407, (2000).

\bibitem{wongphd} W. W. Wong, PhD thesis, University of British Columbia (Canada), arXiv:0811.2826 [astro-ph], (2008).

\bibitem{korobov1} V. I. Korobov, Phys. Rev. A {\bf 61}, 064503, (2000).

\bibitem{drakebook} G. W. F. Drake, {\it Atomic, Molecular and Optical Physics Handbook}, AIP Press, New York, (1996).

\bibitem{kor_monk} V. I. Korobov, D. Bakalov and H. J. Monkhorst, Phys. Rev. A {\bf 59}, R919-R921, (1999).

\bibitem{zheng} X. Zheng, Y. R. Sun, J.-J. Chen, W. Jiang, K. Pachucki and S.-M. Hu, Phys. Rev. Lett. {\bf 118}, 063001 (2017).

\bibitem{yer1} V. Patkos, V. A. Yerokhin, K. Pachucki, Phys. Rev. A {\bf 95}, 012508, (2017).

\bibitem{lu} Z.-T. Lu, P. Mueller, G. W. F. Drake, W. Nrtershuser, Steven C. Pieper, and Z.-C. Yan, Rev. Mod. Phys. 85, 1383 (2013).

\bibitem{gallagher} W. E. Cooke and T. F. Gallagher, Phys. Rev. A 21, 588 (1980).


\bibitem{escobedo1} M. A. Escobedo and J. Soto, Phys. Rev. A {\bf 78}, 032520 (2008).

\bibitem{escobedo2} M. A. Escobedo and J. Soto, Phys. Rev. A {\bf 82}, 042506 (2010).

\bibitem{varsh} D. A. Varshalovich, A. N. Moskalev and V. K. Khersonskii, {\it Quantum Theory of Angular Momentum}, World Scientific, Singapore, (1988).

\bibitem{landau} V. B. Berestetskii, E. M. Lifshitz, and L. P. Pitaevskii, {\it Quantum Electrodynamics}, Pergamon, Oxford, (1982).

\bibitem{Ans} Ya. I. Azimov, A. A. Ansel’m, A. N. Moskalev and R. M. Ryndin, Zh. Exsp. Teor. Fiz. {\bf 67}, 17 (1974) [Sov. Phys.-JETP {\bf 40}, 8 (1975)].


\bibitem{sslg} D. Solovyev, V. Sharipov, L. Labzowsky, and G. Plunien, J. Phys. B: At., Mol. Opt. Phys. 43, 074005 (2010).

\bibitem{derivianko} A. Derevianko and W. R. Johnson, Phys. Rev. A {\bf 56}, 1288 (1997).

\bibitem{bely} O. Bely, J. Phys. B: Atom. Molec. Phys. {\bf 1}, 718 (1968).

\bibitem{drake1} G. W. F. Drake, G. A. Victor, and A. Dalgarno
Phys. Rev. {\bf 180}, 25, (1969).

\bibitem{drake} G. W. F. Drake, Phys. Rev. A {\bf 18}, 820 (1978).

\bibitem{breit} G. Breit and E. Teller, Astrophys. J. 91, 215 (1940).

\bibitem{drake_tab} G. W. F. Drake and Donald C. Morton 2007 ApJS {\bf 170}, 251, (2007).

\end{thebibliography}
\end{document}